\documentclass[pra,twocolumn,showpacs,floatfix]{revtex4}
\usepackage{graphicx}
\begin{document}

\title{Solitary waves and yrast states in Bose-Einstein condensed 
gases of atoms}
\author{A. D. Jackson$^1$, J. Smyrnakis$^2$, M. Magiropoulos$^2$, 
and G. M. Kavoulakis$^2$}
\affiliation{$^1$The Niels Bohr International Academy, The Niels
Bohr Institute, Blegdamsvej 17, DK-2100, Copenhagen \O, Denmark 
\\ $^2$Technological Educational Institute of Crete, P.O. 
Box 1939, GR-71004, Heraklion, Greece}
\date{\today}

\begin{abstract}

Considering a Bose-Einstein condensed gas confined in one 
dimension with periodic boundary conditions, we demonstrate 
that, very generally, solitary-wave and rotational excitations 
coincide. This exact equivalence allows us to establish connections 
between a number of effects that are present in these two problems, 
many of which have been studied using the mean-field approximation.

\end{abstract}
 \maketitle

{\it Introduction.}
Remarkable experimental developments in the field of cold 
atomic gases now permit the realization of systems not previously 
accessible. One example is the recent fabrication of toroidal traps 
that realize periodic boundary conditions \cite{Kurn,Olson,wvk,Hen,Phillips}.  
Tight, elongated traps now permit the study of quasi-one-dimensional 
motion since the energy associated with transverse excitations is much 
higher than the interaction energy \cite{Ket}. A very tight toroidal trap 
is expected soon and will permit the study of periodic motion in a  
quasi-one-dimensional system.

Such advances will make it possible to study many of the 
interesting phenomena predicted by the various nonlinear models 
of such systems that have been studied in recent decades. In the 
case of bosonic atoms, for example, even the ground state is 
nontrivial. When the effective interaction is attractive, a localized 
density maximum forms for sufficiently strong coupling \cite{Ueda,Kav}. 
When the effective interaction is weakly repulsive, the density 
is homogeneous. As the interaction strength increases, however, 
the gas eventually enters the Tonks-Girardeau state of hard-core bosons, 
in which the bosons resemble fermions in many respects \cite{Tonks,Gir}. 

Another interesting aspect of bosonic atoms confined by 
a ring potential is their excitation spectrum \cite{Ts,ZS,Lieb}.  
Two fundamental forms of excitations, solitary-wave excitations 
and rotational excitations, have been investigated.  In the first 
case, one is interested in traveling-wave solutions for which the 
wave propagates around the ring with a constant velocity and 
without change of form. In the second case, one is interested
in the lowest-energy state of the system for a given  
angular momentum, which is the so-called ``yrast" state.

Clearly, both are excited states of the system, and it is
interesting to investigate if and how they are related.  
Bloch's theorem \cite{Bloch} implies that states of 
different values of the winding number are connected via 
excitation of the center-of-mass. This implies that the 
dispersion relation, i.e., the energy of the system as 
function of the angular momentum, is a periodic function 
on top of a parabolic envelope function due to center-of-mass 
excitations. The yrast state is (by definition) the lowest-energy 
state for a fixed angular momentum. Therefore, the naive 
expectation is that the dispersion relation appropriate for 
solitonic excitation should have a structure similar to that 
of the yrast spectrum, but with a higher energy. However, 
in a recent study \cite{KCU}, Kanamoto, Carr, and Ueda have 
provided numerical evidence that ``quantum solitons in the 
Lieb-Liniger Hamiltonian are precisely the yrast states". 
The main result of the present study is to demonstrate 
analytically that solitonic and yrast states are identical.   

The present study contributes to the long-standing problem
of the excitation spectrum of a one-dimensional Bose gas 
with periodic boundary conditions. In his seminal work 
Lieb \cite{Lieb} demonstrated that this spectrum has 
two branches. The usual Bogoliubov spectrum which is 
linear for long wavelengths and quadratic (i.e., 
single-particle like) for higher energies. It is shown here 
that the other branch, which has been identified as corresponding 
to solitary-wave excitations \cite{Kul, Ish}, can also be 
regarded as the lowest-energy state of the gas for a fixed 
value of the angular momentum. 

In the following we first demonstrate the equivalence of 
yrast states and soliton states for the case of an axially-symmetric 
ring. This result is completely general and makes no assumption 
about the explicit form of the many-body state. We then examine 
the limit of weak interactions, where the mean-field 
approximation is an excellent description of both states, 
and discuss the links between various effects that appear 
in these two seemingly different problems.

{\it The equivalence of rotational and solitonic excitations.}
Let us start with the yrast states. For a given Hamiltonian, 
$\hat{H}$, the constraints of a fixed particle number and a 
fixed angular momentum can be imposed with the introduction 
of two Lagrange multipliers, $\mu$ and $\Omega$, which can 
be interpreted physically as the chemical potential and the angular 
velocity of the trapping potential. (We assume here that
the gas has equilibrated in the rotating trap). If 
$\Psi_y(x_1, x_2, \dots, x_N)$ is the yrast many-body 
state, variations of the energy functional,
\vspace*{-1ex}
\begin{eqnarray}
E(\Psi_y, \Psi_y^*) =  
 \int \Psi_y^* \hat{H} \Psi_y \, dx_1 \dots dx_N \ \ \ \ \ \ \ \ 
\nonumber \\
 - \mu \int \Psi_y^* \Psi_y \, dx_1 \dots dx_N
  - \Omega \int \Psi_y^* \hat{L} \Psi_y \, dx_1 \dots dx_N,
\end{eqnarray}
with respect to $\Psi_y^*$ yield 
\begin{eqnarray}
  \hat{H} \Psi_y - \mu \Psi_y - \Omega \hat{L} \Psi_y = 0.
\label{yrequ}
\end{eqnarray}

Turning to the solitonic solutions, we assume that these 
have the form 
\begin{eqnarray}
  \Phi_s(x_1, x_2, \dots, x_N, t) = \Psi_s(z_1,
  z_2, \dots, z_N) \, e^{-i \mu t/\hbar},
\end{eqnarray}
where $z_i = x_i - ut$, where $u$ is the velocity 
of propagation.  Since $ i \hbar {\partial \Phi_s}/
{\partial t} = \hat{H} \Phi_s$,
\begin{eqnarray}
  \hat{H} \Psi_s - \mu \Psi_s - u \hat{P} \Psi_s = 0,
\label{solequ}
\end{eqnarray}
where $\hat{P}$ is the momentum operator. Comparison of 
Eqs.\,(\ref{yrequ}) and (\ref{solequ}) reveals that $\Psi_s$ 
and $\Psi_y$ satisfy the same differential equation and the 
same boundary conditions. They are thus identical with the 
obvious equality $u = \Omega R$, where $R$ is the radius
of the ring.

This equivalence is extremely general and requires only that 
the Hamiltonian is invariant under the transformation $x_i \to 
z_i$. While the kinetic energy and the two-body interaction 
evidently meet this requirement, this is not always the case 
in the presence of an external one-body potential, $V(x)$.  
The above results apply when $V(x)$ is axially symmetric and 
time-independent. The imposition of strict axial symmetry 
has important effects on the exact wave functions $\Psi_y$ and 
$\Psi_s$. It forces them to have an axially-symmetric 
single-particle density distribution, and it implies that the 
condensate is, in general, fragmented. However, these 
conclusions do not necessarily reflect the behaviour of real 
physical systems. In practice, it is impossible to avoid 
weak anisotropies in the trapping potential. Such 
symmetry-breaking anisotropies, even those which are 
vanishingly small in the limit of large atom numbers, are 
sufficient to break the axial symmetry of the single-particle 
density and to restore an unfragmented condensate 
\cite{Rokhsar, ULe, Ho, Liu, Alon, JKM}. 

In the following, we focus on the mean-field approximation, 
which is valid for sufficiently weak interactions. At mean-field 
level, we seek yrast states with a constrained average value of 
$\hat{L}$, rather than insisting on having eigenstates of $\hat{L}$.  
It should be noted that although the yrast state and its
corresponding soliton state are rotationally symmetric, this
rotational symmetry breaks within the mean-field approximation
for both of them. It has been shown, however, in the limit of 
weak interactions that the energies of the axially-symmetric 
and broken-symmetry yrast states are identical to leading 
order in $N$, with differences of order $1/N$ \cite{JKMR, LSY}. 
In the case of solitonic solutions we seek travelling-wave 
solutions, which propagate with a constant velocity. As 
shown more generally above, both problems reduce to the 
same differential equation for the order parameters, with 
the result that $\psi_y = \psi_s = \psi$. Specifically, 
\begin{eqnarray}
 - i \hbar u \frac {\partial \psi} {\partial x} =
- \frac {\hbar^2} {2 M} \frac {\partial^2 \psi} 
{\partial x^2} + (U_0 |\psi|^2 - \mu) \psi. 
\label{yrequ2}
\end{eqnarray}
In this equation we have assumed contact interactions, 
where $U_0$ is the matrix element for zero-energy elastic 
atom-atom collisions (in one dimension), and where $u = 
\Omega R$ as above. We now examine the consequences 
of this equivalence. 

{\it Consequences of the equivalence.}
The solitonic solutions for a Bose-Einstein condensate confined 
to a ring potential with a repulsive effective interatomic 
interaction, $U_0 > 0$, are known to be Jacobi elliptic functions 
\cite{Carr,SMKJ} with density
\begin{eqnarray}
 n(z) = n_{\rm min} + (n_{\rm max} - n_{\rm min}) 
\, {\rm sn}^2 \left( \frac {K(m) z} {\pi R} \Large{|} m \right),
\label{solution}
\end{eqnarray}
where ${\rm sn}(x|m)$ are the Jacobi elliptic functions,
$n_{\rm min}$ and $n_{\rm max}$ are the minimum and maximum
values of the density, and $K(m)$ is the elliptic integral 
of the first kind. The parameter $m$ is determined by the equation
\begin{eqnarray}
 K(m) = \frac {\pi} {\sqrt 2} \frac R {\xi_0} 
\left(\frac {1 - \lambda} m \right)^{1/2},
\label{ellint}
\end{eqnarray}
where $\xi_0$ is the coherence length corresponding to a density 
$n_{\rm max}$, $\hbar^2/(2 M \xi_0^2) = n_{\rm max} U_0$, and 
$\lambda = n_{\rm min}/n_{\rm max}$. The velocity of
propagation is given by
\begin{eqnarray}
 \frac u c = \frac {\sqrt 2 q \xi_0} {R} 
 \pm \frac 1 {2 \pi R} \left[ \lambda 
+ \frac {1- \lambda} m \right]^{1/2} 
\int_{-\pi R}^{\pi R} \frac {\sqrt{n_{\rm min} n_{\rm max}}} 
{n(z)} \, dz.
\nonumber \\
\label{velprof}
\end{eqnarray}
Here $c$ is the speed of sound in a homogeneous gas of density 
$n_{\rm max}$, $M c^2 = n_{\rm max} U_0$, and $q$ is the
phase winding number. According to the above argument, the
yrast states in mean-field approximation are given exactly by 
the same expressions. 

{\it Dispersion.}
We turn now to the dispersion relation, i.e., the energy 
as function of momentum, appropriate for both solitons and 
yrast states. Given the solitary-wave order parameter in 
the form $\psi_s = \sqrt{n} \, e^{i \phi}$, one can determine 
the energy and momentum per particle according to the following 
formulas:
\begin{eqnarray}
  E = \frac 1 {N} \int_{-\pi R}^{\pi R} \left[ \frac {\hbar^2} {2 M}
\left( \left(\frac {\partial \sqrt n} {\partial z} \right)^2 
+ n \left( \frac {\partial \phi} {\partial z} \right)^2 \right)
+ \frac {U_0} 2 \, n^2 \right] d z,
\nonumber \\
\end{eqnarray}
and
\begin{eqnarray}
 p = (\hbar/N) \int_{-\pi R}^{\pi R} n \, ({\partial \phi} 
/{\partial z}) \, dz.
\end{eqnarray}
The analytic form of $\phi(z)$, which is somewhat complicated, 
is given in Ref.\,\cite{SMKJ}. The desired $E(p)$ follows 
immediately given $E(u)$ and $p(u)$. Figure 1 shows the 
results of this calculation for two different situations. Only 
the first two branches of this function are shown. Note that 
$E(-p) = E(p)$. Clearly, the dispersion relation has the 
structure expected from Bloch's theorem as stated above.

It is convenient to introduce the dimensionless quantity 
$\gamma = N U_0 M R/(\pi \hbar^2)$, which is the ratio between 
the interaction energy $n_0 U_0$, with $n_0$ being the homogeneous 
density $n_0 = N/(2 \pi R)$, and the kinetic energy, $E_0 = 
\hbar^2/(2 M R^2)$. In Fig.\,1 the value of $\gamma$ has been 
chosen to be $120/\pi^2$ in the upper curve, and $3/(4 \pi^2)$ 
in the lower curve. In this plot the energy is measured in units 
of $E_0 = \hbar^2/(2 M R^2)$, and the momentum is measured in 
units of $p_0 = \hbar/R$. When $p=0$, the energy $E(p=0)$ comes 
purely from the interaction energy, which is $U_0 N/(4 \pi R)$ 
or $\gamma/2$ in units of $E_0$.  The maximum value of the 
momentum per atom in each ``quasi-periodic" interval is $2 \pi p_0$.  
Finally, the energy difference $E(p = 2 \pi p_0) - E(p = 0) = E_0$, 
and $E(p = 4 \pi p_0) - E(p = 2 \pi p_0) = (2^2 - 1) E_0$. 
More generally, these energy differences are all the odd multiples 
of $E_0$, due to the excitation of the center of mass, as expected 
from Bloch's theorem. 

In the physics of solitary waves, it is well known that the velocity 
of propagation, $u$, of a solitary wave satisfies the equation $u = 
\partial E/\partial p$. This equality can also be viewed from the 
point of view of the yrast states in mean-field approximation. 
Assume that $E(l)$ is the yrast energy per particle as function 
of the angular momentum per particle $l$ and that $E'$ is the energy 
in a rotating frame of reference for which the density distribution 
of the yrast state is stationary. Given that $E' = E - l \Omega$ and 
that $\partial E'/\partial l = 0$, it follows that $\Omega = \partial 
E/\partial l$ or $u = \partial E/\partial p$, which is the formula 
given above. 

\begin{figure}[t]
\centerline{\includegraphics[width=7cm,clip]{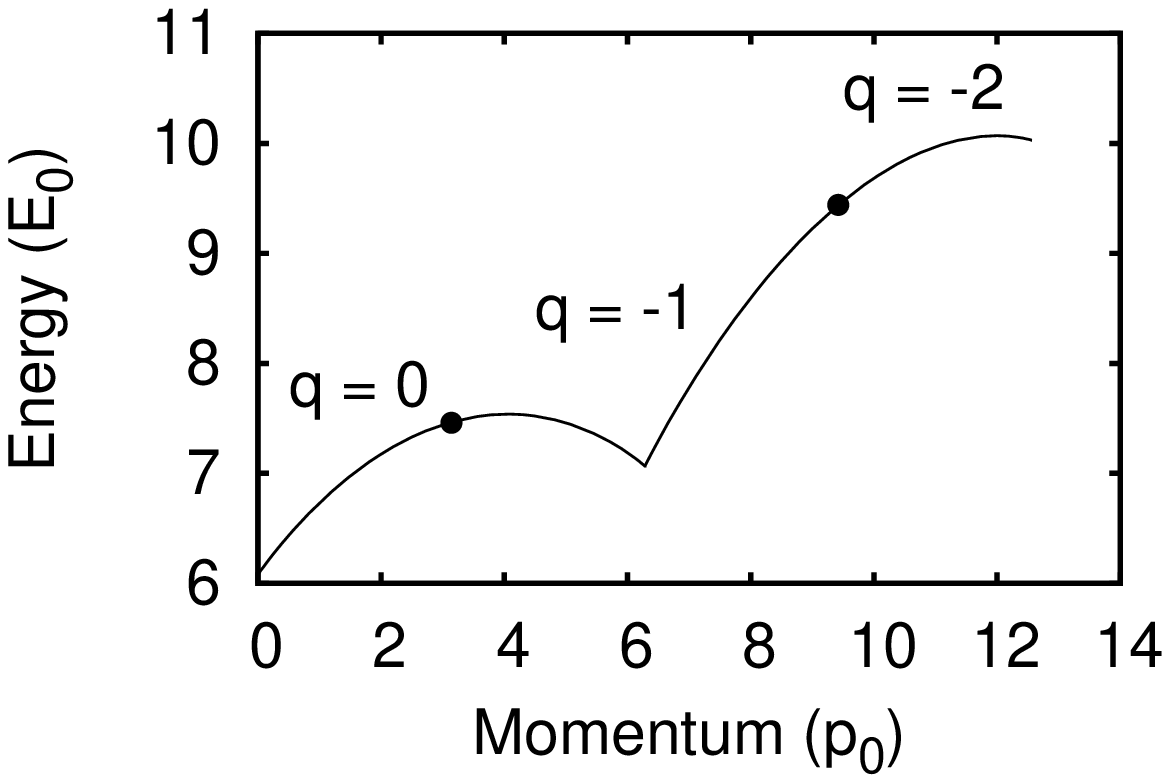}}
\centerline{\includegraphics[width=7cm,clip]{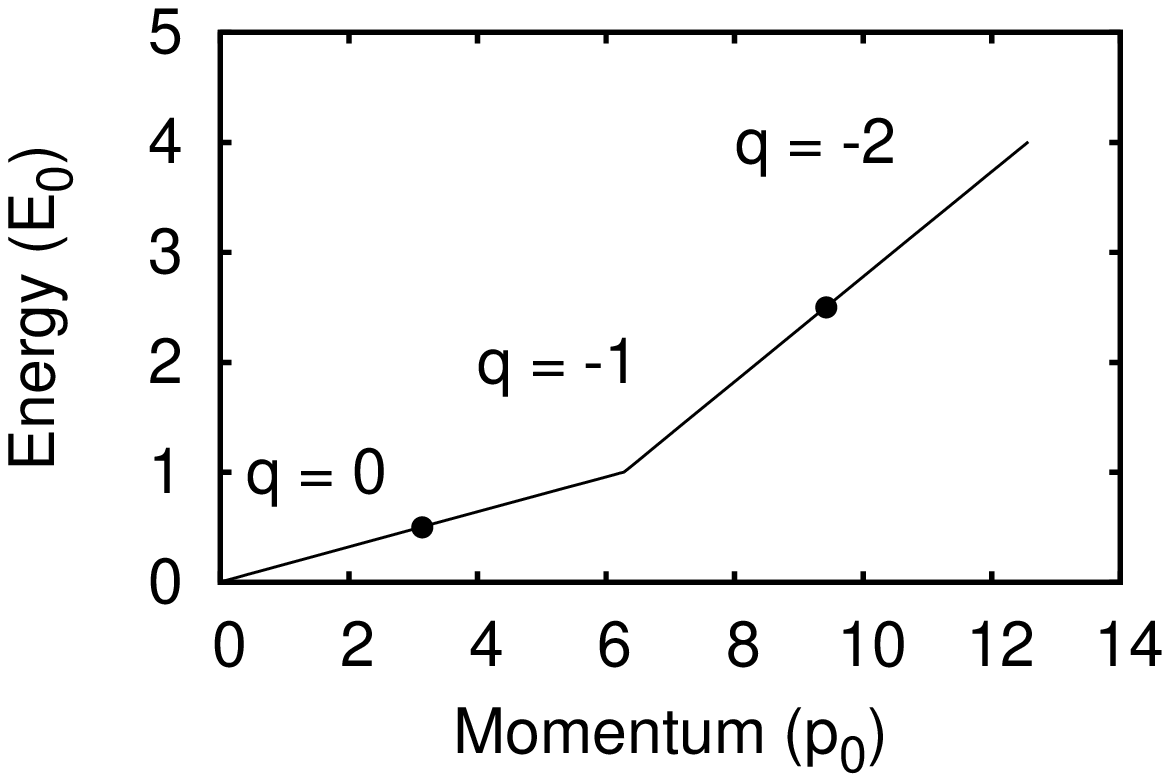}}
\caption{The dispersion relation $E(p)$, for $\gamma 
= 120/\pi^2$ (upper plot), and $\gamma = 3/(4 \pi^2)$
(lower plot). The energy is measured in units of $E_0 
= \hbar^2/(2 M R^2)$, and the momentum in units of $p_0 
= \hbar/R$. In the two curves both $N/R$ and $U_0$ have 
the same value, however in the lower curve the radius
is 40 times smaller than in the higher one.}
\label{fig1}
\end{figure} 

In Ref.\,\cite{SMKJ} it was argued that, when the winding 
number $q$ is zero, the velocity of propagation of a 
solitary wave cannot vanish. This is obvious from the 
dispersion relations of Fig.\,1. The region from $p = 0$ 
to $p= \pi p_0$ (given by the left solid points in the 
two plots of Fig.\,1) corresponds to $q=0$. The smallest 
possible value of $u$ (for $q = 0)$ is attained at $p = 
\pi p_0$ where its value is $u = \hbar/(2MR)$. At this 
value of $p$\, the density develops a node, and a ``dark" 
solitary wave forms.  The density is relatively insensitive 
to changes in the momentum in the vicinity of this point, but 
there is a violent and discontinuous change in the phase of 
the order parameter that is associated with the change of the 
winding number. More generally, the solitary waves become 
dark and the winding number changes whenever $p$ is an 
odd multiple of $\pi p_0$. The first two such points, where 
$q$ changes from 0 to $-1$ and from $-1$ to $-2$, are given 
as solid points in the two plots of Fig.\,1. 

It is also interesting to note that the interaction energy 
(which scales like $N/R$) decreases relative to the zero-point 
energy [which scales like $\hbar^2/(M R^2$)] as the radius of 
the ring decreases for a fixed value of  $N/R$. The ratio of 
these energies is the parameter $\gamma$, which is $\propto 
(N/R) R^2$.  As $R$ decreases for fixed $N/R$, the interaction 
energy becomes less important, and the gas approaches the limit 
of a non-interacting system. In the limit of small $R$, (i.e., 
$\gamma \ll 1$), the quasi-periodic intervals of the dispersion 
relation become linear with a discontinuous first derivative 
between them. The velocity of propagation of the solitary wave 
thus approaches constant values, given as odd multiples of 
$u = \hbar/(2 M R)$, set by the non-interacting problem \cite{SMKJ}. 
This effect is clearly seen in the lower plot of Fig.\,1. 

From the above remarks it is obvious that, as $R$ 
decreases, there is a minimum value of the radius below 
which $u$ cannot vanish \cite{Carr}. The condition for 
this is $R/\xi = \sqrt 6/2$, where $\xi$ is the 
coherence length corresponding to a density $n_0 = 
N/(2 \pi R)$. In the language of yrast states, this says 
that there is a critical coupling of $\gamma = 3/2$
for having a metastable minimum at $L/N = l = \hbar$ 
(or equivalently at $p = 2 \pi p_0$ \cite{Ueda,Kav}. 
For smaller values of $\gamma$, the slope of the dispersion 
relation is positive as $l \to \hbar^-$ [or $p \to (2 \pi p_0)^-$], 
and the possible metastable minimum at this point disappears. 

{\it Energetic stability of the solitary waves.}
The creation of solitary waves in ring-like potentials will 
be of considerable interest. The intimate connection between 
soliton and yrast states discussed here enables us to draw a 
number of conclusions about the possible generation of solitary 
waves and their energetic stability. The most obvious of these 
is that the lowest-energy state with a given $\langle L \rangle$ 
will also be a solitary-wave state. Ideally, one would like to 
observe a family of  solitary-wave solutions much like those 
observed in harmonic traps \cite{Han,Phil}, i.e., density depressions 
travelling with a velocity which is less than that of sound and 
which decreases as the depression becomes deeper.

In the case of a repulsive effective interatomic interaction, 
the curvature of the dispersion relation is negative (as shown 
also in Fig.\,1) \cite{Leg}. The ``kink" states with a momentum 
per particle which is equal to an integer multiple of $2 \pi p_0$ 
are the only possible local minima in the absence of rotation 
of the trap.  The occupation of these states will give rise to 
persistent currents. Since these states have a homogeneous 
density distribution and a non-zero circulation, they would 
be difficult to observe experimentally \cite{rem}. The same 
conclusion applies to a rotating trap. One way to overcome 
this difficulty would be to consider systems with attractive 
effective interactions \cite{KCU}. In this case the curvature 
of the dispersion relation is positive \cite{Leg}, and 
persistent currents cannot be stable. In contrast, solitary 
waves can be stabilized by rotating the trap. The energy in 
a frame rotating with angular velocity $\Omega$ is $E' = E - 
l \Omega$, and $E'$ can have local minima for a continuous 
range of $l$ which is determined by $\Omega$. 

{\it Summary.}
In the present study we have presented a unified picture of 
two seemingly different forms of excitation of a Bose gas 
moving in one dimension with periodic boundary conditions, 
namely rotational and solitonic excitation. We have shown 
that yrast states and solitonic states are identical whenever 
the Hamiltonian is axially-symmetric. This result is a simple 
consequence of the more general observation that the problem 
of finding wave functions that propagate with a fixed velocity 
is equivalent to the problem of minimizing the energy subject 
to a constraint of fixed angular momentum. The only necessary 
condition for this equivalence, which holds for all functional 
forms of the wave function, is the rotational invariance of the 
Hamiltonian. This equivalence permits the consideration of such 
systems from two distinct points of view, and we have noted 
certain aspects of a unified picture that result from this 
observation in the mean-field regime, which is valid for 
sufficiently weak interactions. 

This connection between solitons and yrast states has been 
suggested previously \cite{KCU} and the arguments offered here 
are admittedly elementary. It would appear, however, that the 
rigorous equivalence demonstrated here is neither generally 
known, nor adequately appreciated.


\begin{thebibliography}{99}
\bibitem{rem} Since a necessary (but not sufficient) condition 
for metastability is that the density distribution is homogeneous, 
while repulsive interactions are necessary to create persistent 
currents, the metastability of superflow works against the 
observation of the solitary waves with an inhomogeneous density
distribution.
\end{thebibliography}
\end{document}